\documentclass[fleqn,usenatbib]{mnras}

\usepackage{newtxtext,newtxmath}

\usepackage[T1]{fontenc}
\usepackage{ae,aecompl}

\usepackage{graphicx}	
\usepackage{amsmath}	
\usepackage{amssymb}	
\usepackage{commath}	

\let\oldhref\href
\renewcommand{\href}[2]{\oldhref{#1}{\hbox{#2}}}

\title[Deconstructing M83]{Deconstructing a galaxy: colour distributions of point sources in Messier 83%
\footnote{  
Based on observations made with the NASA/ESA Hubble Space Telescope, obtained from the Data Archive at the Space Telescope Science Institute, which is operated by the Association of Universities for Research in Astronomy, Inc., under NASA contract NAS 5-26555. These observations are associated with program \#11360.
}
}
\author[Kiar, Barmby \& Hidalgo]
{
A. K. Kiar$^{1}$\thanks{E-mail: akiar@uwo.ca},
P. Barmby$^{1}$\thanks{E-mail: pbarmby@uwo.ca} and
A. Hidalgo$^{1,2}$\thanks{E-mail: andrea@inaoep.mx} \\
$^{1}$Department of Physics and Astronomy and Centre for Planetary Science and Exploration,
University of Western Ontario, London, ON, N6A 3K7, Canada\\
$^{2}$Instituto Nacional de Astrof\'{i}sica, \'{O}ptica y Electr\'{o}nica, Luis Enrique Erro \# 1, Tonantzintla, Puebla, Mexico\\
}

\pubyear{2017}

\begin{document}
\maketitle

\begin{abstract}
What do we see when we look at a nearby, well-resolved galaxy?
Thousands of individual sources are detected in multi-band imaging observations of even a fraction of a nearby galaxy,
and characterizing those sources is a complex process.
This work analyzes a ten-band photometric catalog of nearly 70~000 point sources in a 7.3 square arcminute region of the nearby spiral galaxy Messier 83,
made as part of the Early Release Science program with the Hubble Space Telescope's Wide Field Camera 3.
Colour distributions were measured for both broad-band and broad-and-narrow-band colours;
colours made from broad bands with large wavelength differences generally had broader distributions although
$B-V$ was an exception.
Two and three dimensional colour spaces were generated using various combinations of four bands and clustered with the K-Means and Mean Shift algorithms.
Neither algorithm was able to consistently segment the colour distributions: while some distinct features in colour space were apparent in visual examinations, 
these features were not compact or isolated enough to be recognized as clusters in colour space.
$K$-Means clustering of the $UBVI$ colour space was able to identify a group of objects more likely to be star clusters.
Mean Shift was successful in identifying outlying groups at the edges of colour distributions.
For identifying objects whose emission is dominated by spectral lines, there was no clear benefit from combining narrow-band photometry in multiple bands compared to a simple continuum subtraction.
The clustering analysis results are used to inform recommendations for future surveys of nearby galaxies.

\end{abstract}


\section{Introduction}
Galaxies are complex systems, comprised of numerous components with enormous ranges of size, mass, density, and composition.
These components can be divided into baryonic (stars and their
remnants, nebulae, star clusters, nuclear black hole) and non-baryonic (dark matter);
detecting the components and describing the interactions between them is a key step in elucidating the natural history of galaxies.
Only in nearby galaxies can individual sub-components be resolved.
As observational technology has advanced, the definition of ``nearby'' has changed and will continue to do so.
Stars can be resolved in Milky Way satellites and Local Group galaxies to distances of about 1~Mpc with the {\em Hubble Space Telescope (HST)};
this limit is expected to increase to about 1.5~Mpc with the James Webb Space Telescope and reach the distance of the nearest large elliptical (3.5~Mpc) with potential future facilities \citep{brown12}.

Subcomponents of galaxies are most often detected via multi-band imaging.
Here we focus on components with effective temperatures in the range $3-10\times 10^4$~K,
detectable in imaging at near-infrared through ultraviolet wavelengths.
Particular stellar types, or star clusters, are often identified with broad-band colour-magnitude diagrams \citep[e.g.][]{chandar10}.
Narrow-band filters can also isolate special stellar types \citep[e.g. Wolf-Rayet stars,][]{hadfield05} or nebulae prominent in emission
lines such as planetary nebulae or supernova remnants \citep{herrmann08,blair14}.
Spectroscopic follow-up is often required to confirm the nature of candidates.
New observational facilities which provide spatially-resolved spectroscopy \citep[e.g.][]{yan16,sanchez12,drissen10} may reduce the need for separate imaging and follow-up steps,
at the cost of increased complexity in the initial data analysis.

Multi-wavelength imaging surveys are very common in studies of unresolved galaxies in the distant universe.
While these are often designed to select galaxies or active galactic nuclei (AGN) with specific properties \citep[e.g.][]{trenti11,timlin16},
sometimes they are pure blank-field surveys.
Broadband ($R=\Delta \lambda/\lambda \lesssim 5$) filters are the most common imaging modality,
although there have been some narrow- or medium-band surveys as well \citep[e.g. ][]{combo-17}.
Clustering in colour space has been used to select particular classes of objects from surveys,
for example AGN  \citep[e.g.][]{secrest2015, dabrusco09} or extragalactic star clusters \citep[e.g.][]{dabrusco16, hollyhead15}.
The advantage of using clustering for identification is that it is {\em data-driven}, relying on observed properties rather than expectations. 
Unusual classes of objects or unexpected effects (such as a non-standard extinction law) are thus more likely to be discovered, if present.
To our knowledge, clustering in colour space has not been used in exploratory data analysis of nearby galaxy imaging.

The purpose of this work is to treat a nearby galaxy as if it were a blind survey field, examining the colour distributions of the detectable point sources and the extent to which they separate into distinct groups in two- and three-dimensional colour spaces.
The dataset used for this study is the Wide-Field Camera-3 (WFC3)
Early Release Science (ERS) observations of the nearby spiral galaxy Messier 83 (M83).
M83 is a grand-design spiral of type SAB, located at a distance of 4.66~Mpc \citep{tully13}
and the largest member of the M83 subgroup of the nearby Centaurus group of galaxies \citep{tully15}.
It has a complex nuclear region \citep{mast06,thatte00} and has hosted 6 supernovae in the past century \citep{stockdale06}.
This study uses the catalogue of M83 point sources produced by \citet{chandar10} from the ERS WFC3 imaging at ultraviolet to near infrared wavelengths (200--900~nm).
We form colours from the photometric measurements in the catalogue, investigate the colour distributions,  and apply several clustering techniques to the resulting multi-dimensional colour  datasets.
We evaluate the utility of different methods and colour combinations for identifying galaxy sub-components.

\section{Data}
\subsection{Imaging dataset}

The WFC3 ERS observations of M83 were made in broad- and narrow-bands in order to characterize both stellar and nebular properties.
They cover a $3.6\times3.6$~kpc$^2$ region in the northern part of the galaxy, including the nucleus,
a portion of a spiral arm and an interarm region.
The galaxy's apparent diameter of $\sim13$~arcmin \citep{rc3} is reasonably well-matched to the camera's field of view.
The spatial resolution of the images is $0\farcs0396$~arcsec~pixel$^{-1}$,
corresponding to a linear scale of $0.9$~pc~pixel$^{-1}$ at the 4.66~Mpc distance.
A complete description of the observations and data processing is given by \citet{chandar10}.
Our work here uses the observations in the UVIS channel, listed in Table~\ref{tab:filters} with filter information from the STScI website.%
\footnote{\url{http://www.stsci.edu/hst/wfc3/ins_performance/ground/components/filters}}
A number of previous studies have used the ERS M83 dataset for various purposes.
These include studies of  star clusters \citep{chandar10, wofford11, whitmore11, bastian11, bastian12, fouesneau12, silva13, andrews14, chandar14, adamo15,ryon15,hollyhead15, sun16},
\ion{H}{ii} regions \citep{liu13}, supernova remnants and the interstellar medium \citep{dopita10, hong11, blair14, blair15}, 
resolved stars \citep{kim12, williams15},
and a super-Eddington off-nuclear black hole \citep{soria14}.

\begin{table}
\centering
\caption{Filters used in M83 ERS survey}
\label{tab:filters}
\begin{tabular}{lllll}
\hline\hline
Band & Name & Peak $\lambda$ & $\Delta \lambda$ (FWHM) & Exp.\ time\\
& & nm & nm & s\\
\hline
F225W &  Wide UV           & 225 & 50.0 & 1800\\
F336W &  $U$-band          & 338 & 55.0 & 1890\\ 
F373N &  [\ion{O}{iii}]    & 373 & 3.8 & 2400\\
F438W &  $B$-band          & 432 & 69.5 & 1180\\
F487N &  H~$\beta$         & 487 & 4.5 & 2700\\
F502N &  [\ion{O}{ii}]     & 501 & 47.0& 2484\\
F555W &  $V$-band          & 541& 160.5 & 1203\\
F657N &  H~$\alpha$+[\ion{N}{ii}]&  657 & 9.4 & 1484\\ 
F673N &  [\ion{S}{ii}]     & 673 & 7.7 & 1850\\
F814W &  $I$-band          & 835 & 255.5 & 1203\\
\hline
\end{tabular}
\end{table}

We analyze the catalogue produced by \citet{chandar10} and made available via the Mikulski Archive for Space Telescopes,\footnote{\url{https://archive.stsci.edu/pub/hlsp/wfc3ers/hlsp_wfc3ers_hst_wfc3_m83_cat_all_v1.txt}}
hereafter referred to as the `ERS catalog.'
The sources in this catalogue were detected on a `white-light' image produced by a weighted combination of the $UBVI$ images.
This detection method is expected to be less biased against very red or blue sources than single-filter detection, although it may still miss objects whose emission is emission-line (rather than continuum) dominated.

Photometry in 0.5- and 3-pixel radius apertures at the positions of the detected sources was performed on the broad- and narrow-band images and tabulated in the Vega magnitude system. 
The catalogue contains about 68~000 sources which are expected to include individual stars in M83, star clusters, stellar blends,
supernova remnants, \ion{H}{ii} regions, planetary nebulae, and background galaxies.
The high Galactic latitude ($b=+32^{\circ}$) of M83 means that foreground star contamination is not expected to be substantial.
Completeness and reliability of the catalogue are not discussed by \citet{chandar10},
but a visual inspection of the the detected sources on the white-light image suggests that a reasonable balance between completeness and reliability was achieved.
Nine sources are flagged in the catalogue as being problematic  and we remove them from our analysis.
We apply the correction to the F657N magnitude zeropoint (from 20.72 to 22.35) noted in the header of the catalog.
\citet{chandar10} discussed aperture corrections for this catalog, but since we are primarily concerned with colours
and the aperture correction does not vary strongly with wavelength, we
omit it.

As a check on the catalogue we used SExtractor to detect and photometer sources in the single-band images.
While the aperture photometry measurements matched well, the derived uncertainties were much smaller than those reported in the catalog.
Indeed, the catalogue uncertainties seem to be physically unreasonable, with median uncertainty values well above 1~magnitude in
most bandpasses, and the catalogue notes do not recommend them for use except in a relative sense.
Our comparison implied that recovering a more typical magnitude uncertainty distribution would be accomplished by dividing the
0.5-pixel magnitude uncertainties by 10 for the broad-bands, 15 for the narrow-bands, and 8 for the F657N (H~$\alpha$) band.
This allows us to use the catalogue aperture magnitudes as an indicator of detected signal-to-noise: our analysis uses only sources with (scaled) 0.5-pixel magnitude uncertainties $<0.2$~mag.
For the remainder of the analysis we use magnitudes measured in the 0.5-pixel radius aperture, as these should be less affected by crowding and the variable galaxy background.

\begin{figure*}
\centering
\includegraphics[width=0.49\textwidth]{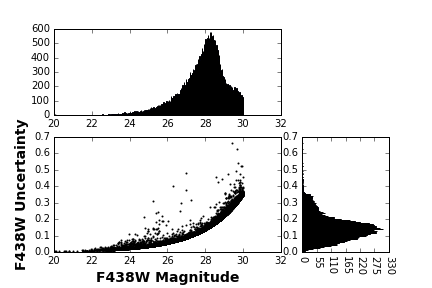}
\hfill
\includegraphics[width=0.49\textwidth]{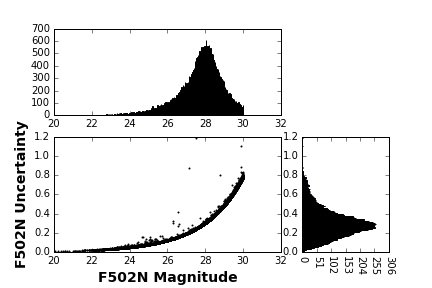}
\caption{Distribution of magnitudes and scaled uncertainties for sources in the \citet{chandar10} M83 ERS catalogue. 
Uncertainties were scaled as outlined in the text. Magnitudes $>30$~mag are not shown.} 
\label{fig:mag_unc}
\end{figure*}

Table~\ref{tab:cat_numbers} and Fig.~\ref{fig:mag_unc} characterize the catalogue in terms of measurements in individual bands.
Only about 9\% of the sources are detected in all bands: 
Table~\ref{tab:cat_numbers} gives the number of sources for which photometry is reported in a given band ($N_{\rm det}$),
the number for which the scaled 0.5-pixel magnitude uncertainty is $0.2$~mag or less ($N_{\rm good}$),
and the aperture magnitude at which the median magnitude uncertainty is $0.2$~mag ($m_{\rm good}$). 
(We remind the reader that aperture corrections have not been applied to these magnitudes.)
Most of the detections in the broad-band images are of sufficient signal-to-noise for reliable photometry, but this is less true for the F225W and narrow-band images, which were not used to construct the detection image.
The photometry is deepest in the F555W band, as expected since it is at the centre of the detection image's wavelength range.
Fig.~\ref{fig:mag_unc} shows the distributions of magnitudes and corresponding uncertainties in example broad and narrow bands. 
This figure illustrates that the majority of sources have a scaled photometric uncertainty $<1$~mag and validates the use of 0.2~mag uncertainty as a detection limit in individual bands. 
This figure also illustrates that the magnitude peak occurs between 25 and 28~mag. 

\begin{table}
\centering
\caption{Catalogue source counts in individual bands}
\label{tab:cat_numbers}
\begin{tabular}{lrrlr}
\hline\hline
band & $N_{\rm det}$ & $N_{\rm good}$ & $m_{\rm good}$ \\
\hline
F225W &  57237 & 15011 & 25.2 \\
F336W &  62192 & 34129 & 26.6 \\
F373N &  55966 & 8878 & 24.8 \\
F438W &  66356 & 48858 & 28.0 \\
F487N &  63812 & 13335 & 25.8 \\
F502N &  64313 & 14654 & 26.4 \\
F555W &  67424 & 65652 & 30.0 \\
F657N &  67782 & 23939 & 26.9 \\ 
F673N &  65305 & 25295 & 26.3 \\
F814W &  67050 & 59600 & 27.9 \\
\hline
\end{tabular}
\end{table}

\subsection{Colour Selection}

The ERS observations in 10 bands allow the generation of 45 different colours, but not all of these colours are likely to be useful in characterizing components of the galaxy.
A major purpose of this work is to explore which four-band combinations are most useful.
Typical observations of nearby galaxies involve three or four bands,  which can be used to construct two and three independent sets of colours, respectively. 
With four bands $ABCD$, colours can be constructed in either two dimensions (e.g., $A-B$ versus $C-D$) or three (e.g., $A-B$ versus $B-C$ versus $B-D$ and other combinations).
Both variations were considered in the clustering analysis.
While two-dimensional colour spaces are more familiar to astronomers and simpler to visualize, they do not fully capture all of the colour information available from four bands. 
Comparing two and three-dimensional colour spaces was another goal of this work.

Two types of colour combinations were created: combinations of the most commonly-used broad bands and combinations including three broad bands and one narrow band.
For three-dimensional colour spaces, a common band between the three colours was used in order to easily generate colours that could be transformed into the original two dimensional space. 
In the broad band combinations, three dimensional colour spaces used either F438W or F555W as the common band.
In the narrow band combinations, the narrow band was used as the common band.
Although the original ERS catalogue contained approximately 68000 sources, not all sources were detected in all bands with sufficient signal-to-noise for reliable colours. 
For a given combination, only sources well-detected in all bands (magnitude uncertainty $<0.2$~mag) were used in the clustering analysis.

\subsubsection{Broad Band Combinations}

The first type of combination was comprised of the broad bands: F336W, F438W, F555W, and F814W.%
\footnote{
For readability, we refer to these hereafter as $U$, $B$, $V$ and $I$, although these bands do not correspond exactly to the ground-based equivalents.}
Broadband 300--800~nm colours are rough indicators of stellar temperatures, reddening, and (indirectly) age and metallicity.
These bands are used in many \textit{HST} studies and had the largest number of detections in the ERS catalog. 

About 33~000 objects had reliable $U - B$ or $U-V$ colors, 41~000 had $B-I$ and nearly $58~000$ had reliable $V-I$ colours.
Although also a broad-band filter, the UV-wide filter F225W (hereafter UVW) is less-commonly used in the literature, and for the purposes of creating colour combinations it was treated as a narrow-band filter.
When creating colours from the broad bands, $U - I$ was not used as it was also not commonly-used in the literature. 

\subsubsection{Narrow Band Combinations}

Imaging of galaxies in narrow bands is typically used to select sources bright in particular emission lines, for example \ion{H}{ii} regions in H~$\alpha$ or planetary nebulae in \ion{O}{ii}[5007].
Since the ERS catalogue was constructed from broad-band imaging, it was unknown at the start of our analysis how many emission-line sources would be included; the clustering analysis with these bands was more exploratory.
The second set of colour combinations included the narrow bands F373N (\ion{O}{ii}), F487N (H~$\beta$), F502N (\ion{O}{iii}), F657N (H~$\alpha$), F673N (\ion{S}{ii}), and the broad band F225W (UVW).
Colours were created  by pairing each narrow band with the broad band which overlapped it in wavelength space.
This was done to separate sources whose spectra are emission line dominated from continuum-dominated sources.
The second colour in each combination was created from two broad bands that did not overlap the first colour in wavelength space.
Table~\ref{tab:NBcolourcombos} lists the narrow band colour combinations used for analysis.
Compared to the broad-band colours, the narrow-band combinations generally contained fewer sources with less dense colour distributions and this provided a different regime in which to test the clustering.
The number of sources in the narrow band combination, with the exception of the $H\alpha$ band, is significantly lower than the broad band combinations. 
Table \ref{tab:NBcolourcombos} lists the two dimensional colour combinations, the number of sources detected in each colour and their mean uncertainty, and the number of sources detected in the colour combination for each narrow-band colour.

\begin{table}
\centering
\caption{Narrow band colour combinations: detections in individual colours and combinations}
\label{tab:NBcolourcombos}
\begin{tabular}{lrl}
\hline\hline
Bands & $N_1^a$ & Mean Unc. \\
\hline
$UVW - U$ &  14977 & 0.15 \\
$B - V$ &  14943 & 0.14  \\
$B - I$ &  14095 & 0.15  \\
$V - I$ &  14098 & 0.13  \\
\hline
$U - {\rm F373N}$ & 8675 & 0.15 \\
$B - V$ &  8657 & 0.15 \\
$B - I$ &  8558 & 0.16 \\
$V - I$ &  8559 & 0.13 \\
\hline
$B - {\rm F438N}$ & 13269 & 0.14 \\
$V - I$ &  13147 & 0.13 \\
\hline
${\rm F502N}- V$ & 14644 & 0.14 \\
$U - B$ &  13390 & 0.16 \\
\hline
${\rm F657N} - I$ & 59465 & 0.14 \\
$U - B$ &  28920 & 0.16 \\
$U - V$ &  28920 & 0.14 \\
$B - V$ &  41317 & 0.14 \\
\hline
${\rm F673N} - I$ & 25185 & 0.15 \\
$U - B$ &  14577 & 0.16 \\
$U - V$ &  14586 & 0.14 \\
$B - V$ &  18882 & 0.14 \\
\hline
\end{tabular}

$^aN$ is number of sources detected with photometric uncertainties $<0.2$~mag.
\end{table}

\subsection{Model colours}

\begin{figure}
\centering
\includegraphics[width=0.5\textwidth]{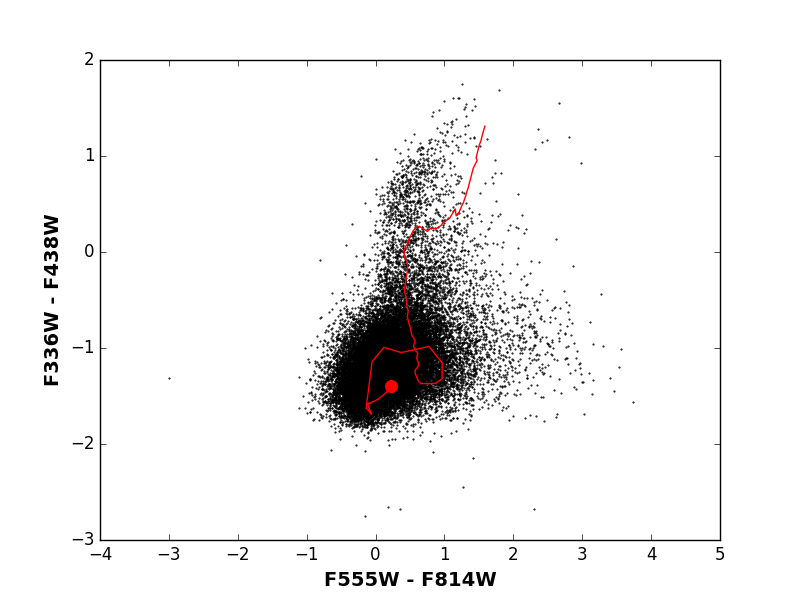}
\caption{ERS catalogue distribution of sources with reliable $V - I$ and $U - B$ colours. FSPS single stellar population model (${\rm Fe/H}=+0.5$, ages $10^5-10^{10.3}$~yr) is over-plotted as a solid line; the large circle on this line indicates the youngest age.}
\label{fig:model_colour}
\end{figure}

As a check on the reasonableness of the colours generated from the ERS catalog, we generated single stellar population model colours using the Flexible Stellar Population Synthesis code \citep[FSPS;][]{conroy09,conroy10}. 
Magnitudes of a stellar population formed in a single burst of star formation were generated using the default FSPS parameters as implemented in Python-FSPS including MIST isochrones and a \citet{kroupa01} initial mass function. 
The only non-default parameter was the inclusion of nebular emission based on a Cloudy model \citep{byler16}. 
Fig.~\ref{fig:model_colour} shows an example colour track in four commonly-used broad bands, for a super-solar-metallicity (${\rm Fe/H}=+0.5$) population.
The colours are bluest at the youngest ages ($10^5$~yr), undergo a loop in colour space for ages $10^{6.8}<t<10^{7.9}$~yr when the first asymptotic giant branch stars become important, and then monotonically redden to the oldest ages. 
The colours of the individual ERS catalogue sources, also plotted in the Figure, are approximately consistent with the track colours, giving confidence that the observed colours are reasonable.

\section{Methods}
Clustering methods provide an efficient way of finding structure in high dimensional data.
Numerous techniques for clustering multi-dimensional data are available.
We used two well-known algorithms, $K$-Means and Mean Shift;
both were implemented using the {\sc sklearn.cluster} Python package version 0.17 \citep{sklearn}. 
We also investigated the use of a newer and less well-known algorithm, affinity propagation \citep{frey07}.
Affinity propagation calculates the similarities between the data points as input for clustering, and uses a series of ``messages'' between data points to determine the number of clusters and their centres.
We found this algorithm to be very sensitive to the input parameters and also rather slow due to the calculation of the messages passed between points on each iteration, and chose not to use it further.

\subsection{K-Means Clustering}
K-Means is one of the most widely used clustering methods. 
In astronomy, K-Means has has been used to analyze a variety of different objects including
ultraviolet quasar spectra \citep{tammour16}, supernova light curves \citep{rubin16}, and structures in stellar phase space \citep{hogg16}.
It is simple, robust, and easy to implement when analyzing high dimensional spaces, making it a powerful way to analyze multi-band photometric surveys. 
The K-Means algorithm requires the number of clusters $K$ to be selected in advance.
The algorithm is initialized by selecting $K$ data points at random and designates these points as cluster centres, denoted by $\mu_k$.
Each of the $n$ points $x_i$ in the  data set is then assigned to a cluster centre by finding the centre to which the distance is the smallest.
K-Means aims to minimize the sum of squares of distances within each cluster given by:

\begin{equation}
\label{eq:km_sos}
J = \sum_{i=1}^n \sum_{k=1}^K \min \big( \norm{x_i - \mu_k}^2 \big)
\end{equation}

where $\norm{x_i - \mu_k}$ indicates the distance measurement in $N$-dimensional space. 
In this work the Euclidean distance  is used, but other distance metrics are also possible.
Each algorithm iteration re-calculates the cluster centres by taking the average position of all the points in each cluster as the new centre. The points are reassigned to the new nearest cluster centre.
The stopping criterion is that the change in centre location is less than a given threshold for two consecutive iterations \citep{sklearn.cluster}. 

The requirement to select the number of clusters in advance is a disadvantage of K-Means.
In high-dimensional data which cannot be easily visualized, determining the number of clusters by inspection is not straightforward. 
A given dataset may not have an optimal number of clusters, but measures of clustering effectiveness can be used to discriminate 
between values.
In order to reduce the uncertainty in determining the number of clusters, we developed a process to identify the behaviour of various clustering parameters,
described in Section~\ref{sec:cluster_process}.

\subsection{Mean Shift Clustering}
\label{sec:methods_ms}

Mean Shift is a non-parametric clustering technique based on probability density function estimates at each point in a multi-dimensional data set. 
Although common in fields such as remote sensing, Mean Shift has not been widely used in astronomy. 
In one of the few examples found, it was used by \citet{gomez10} to find structures in the energy-angular momentum space occupied by particles in an $N$-body simulation.
The power of Mean Shift clustering is that the clusters it creates are not confined to a particular shape.
Because Mean Shift moves towards the local mode near the data on which it was initialized, it is useful for estimating the number of clusters in the data \citep{comanciciu02}.
At each point, the algorithm estimates the density around that point using a small sample of nearby objects.
The algorithm is based on two components: kernel density estimation and density gradient estimation.
The following highlights the major components of the algorithm; for a full description, see \citet{vatturi09}.

In clustering $x_i$, a set of $n$ independent $d$-dimensional data points, the first element of Mean Shift is kernel density estimation.
The density estimator for a multivariate density kernel $K_{\textbf{H}}(x) = \abs{\textbf{H}}^{0.5} K(\textbf{H}^{0.5} x)$ is given by \citet{vatturi09}: 
\begin{equation} 
\label{eq:ms_de}
\hat{f}(x) = \frac{c_{k,d}}{nh^d} \sum_{i=1}^n k\Bigg( \norm{\frac{x-x_{i}}{h}}^2 \Bigg)
\end{equation}
where  $c_{k,d}$ is a normalization constant and $k(x)$ satisfies
\begin{equation}
K(x) = c_{k,d} k(\norm{x}^2).
\end{equation}
The $K$ defined for the kernel should not be confused with the number of clusters $K$ defined for the K-Means algorithm. 

The major parameter of Mean Shift is bandwidth, $h>0$, which appears in the bandwidth matrix $\textbf{\textit{H}} = h^2\textbf{\textit{I}}$.
Estimating bandwidth correctly is critical to determining the correct number of clusters.
If the bandwidth is too low, the density estimate will be under-smoothed, and Mean Shift will produce many small clusters.
Conversely, if the bandwidth is too large, a small number of large clusters will be produced, resulting in groupings of data that may blur the underlying structure \citep{vatturi09}.

In order to determine the correct bandwidth for each clustering, the {\sc estimate-bandwidth} function from {\sc sklearn-cluster} was used. This function estimated the bandwidth by calculating the variance between points in the data, and used the minimum value as the bandwidth. 

The second element of Mean Shift is density gradient estimation. 
The density gradient is estimated from the gradient of equation~\ref{eq:ms_de} and given by: 
\begin{multline}
\label{eq:ms_dg}
\nabla\hat{f}_{h,K}(x) = \frac{2c_{k,d}}{nh^{(d+2)}} \Bigg[ \sum_{i=1}^n k^{\prime} \Bigg( \norm{\frac{x-x_{i}}{h}}^2 \Bigg) \Bigg] \\ \Bigg[\frac{\sum_{i=1}^n x_{i}k^{\prime} \Bigg( \norm{\frac{x-x_{i}}{h}}^2 \Bigg)}{\sum_{i=1}^n k^{\prime} \Bigg( \norm{\frac{x-x_{i}}{h}}^2 \Bigg)} - x \Bigg]
\end{multline}

The second term of equation~\ref{eq:ms_dg} is the Mean Shift, the difference between the weighted mean using $k^{\prime}$, the derivative of the profile of the kernel $K(x)$, and $x$.
Applying a Gaussian kernel, the Mean Shift $m_{h,K}(x)$  becomes: 
\begin{equation} 
\label{eq:ms_sh}
m_{h,K}(x) = \frac{\sum_{i=1}^n x_{i} \exp \Bigg(\norm{\frac{x-x_{i}}{h}}^2 \Bigg)}{\sum_{i=1}^n \exp \Bigg(\norm{\frac{x-x_{i}}{h}}^2 \Bigg)} - x
\end{equation}
The Mean Shift always points in the direction of largest ascent through the estimated density function \citep{vatturi09}, causing the algorithm to converge to areas of high density. 

Mean Shift clustering involves the iterative application of equation~\ref{eq:ms_sh} to shift the points of a data set towards the direction of the Mean Shift vector.
In each iteration $j$ the points are shifted by: 
\begin{equation}
\label{eq:ms}
x_i^{j+1} = x_i^j + m_{h,K}(x_i^j)
\end{equation}
The {\sc sklearn.cluster} implementation of Mean Shift stops when the shift is $<10^{-3}h$ or a maximum number of iterations is reached.
Shifting the data points via equation~\ref{eq:ms} ensures that when the points converge, the center is the area of highest local density. 
The density mode can be interpreted as the centre of a significant cluster in the data set and is used to classify the points shifted towards it.

Mean Shift clustering is prone to selecting one large cluster surrounded by several small clusters containing only $1-2$\% of the total number of data points. 
This is because the algorithm is drawn to areas of high density, which causes it to assign a large volume of data points to one cluster. 
Mean Shift is also very sensitive to bandwidth selection, and results vary drastically based on the bandwidth parameter.

\subsection{Clustering Process}
\label{sec:cluster_process}

\subsubsection{Mean Shift}

Mean Shift clustering was performed first by estimating the bandwidth parameter with the {\sc estimate-bandwidth} function in the Python {\sc scikit-learn} package \citep{sklearn}.
This function estimates the bandwidth parameter based on the distances between points in the data, and determines if the distribution has high or low variance. 
Following the initial clustering, the bandwidth was varied and the clustering performed again to determine how sensitive a combination was to the parameter. 
The bandwidth values were changed by increments of $\pm 0.1$ or $\pm 0.05$ from the estimated bandwidth value depending on a combination's sensitivity to the parameter. 
If a combination was very sensitive to bandwidth, then the number of clusters found by Mean Shift would vary greatly over a small range of bandwidth values.
This type of combination usually resulted in poor segmentation, as the algorithm would not converge on a number of clusters. 
However, sensitivity could also be the result of the starting bandwidth estimate.
If the original estimate was in an unstable bandwidth interval, it would be reflected in the bandwidth hierarchy. 
The testing of multiple bandwidth values was expected to result in convergence of the number of clusters. 

\subsubsection{K-Means}

K-Means clustering was performed after Mean Shift, allowing us to use the number of clusters determined by Mean Shift $K_{\rm MS}$ as an initial estimate.
Next, K-Means was performed with $K_{\rm MS}-3\leq K \leq K_{\rm MS} +3$  
to explore the algorithm performance with different values of $K$.
We found that, compared to Mean Shift, K-Means converged more quickly and tended to produce clusters closer in size to one another.

Several checks of clustering reliability were made.
For each K-Means clustering, the sum-of-squares value versus $K$ was plotted: 
the sum-of-squares represents the distance between every point within a cluster.
It was expected that as $K$ increased and there were fewer sources in each cluster, the sum-of-squares would decrease, and this was found to be the case.
Following the sum-of-squares test, we tested the reproducibility of the clusters produced by the K-Means algorithm. 
Since K-Means is initialized randomly, the clusters produced can depend on the starting position.
The cluster centres were tested by running K-Means for 40 trials with the same value of $K$: while the initial cluster centres were different, the final cluster centres were the same to within $\pm 0.1$ magnitude.

\subsection{Clustering Statistics}

The following method allowed the investigation of the effect of input parameters on each technique.
This process identified the clustering which was most successful at identifying different segments of sources in the colour space.

Selecting the optimal clustering for a data set can often seem arbitrary, as no ``correct" answer necessarily exists.
In order to characterize the clustering results, a variety of metrics were calculated.
The relationships between clustering parameters were investigated to determine how they indicated the strength of clustering.
Since the performance of the algorithms was directly related to the input parameters, those relationships were critical for characterizing the clustering.
Astrophysical interpretation of cluster membership also plays an important role in understanding the utility of an algorithm/dataset combination; this is discussed further in Sect.~\ref{sect:results}.

Various statistics were calculated to describe cluster properties.
The average colour and standard deviation were calculated for each cluster in order to describe the distribution of the sources in each cluster in the colour space. 
The fractional size of each cluster (relative to the entire dataset) was calculated to describe the distribution of sources between clusters.
In addition to descriptive statistics, a clustering metric was also used to characterize each clustering. The silhouette score is a metric used to describe cluster compactness and isolation.
The silhouette score is given by:
\begin{equation}
\label{eq:ss}
S = \frac{1}{N} \sum{\frac{b - a}{\max\big(a, b\big)}}
\end{equation}
where $a$ is the mean intra-cluster distance, and $b$ is the distance between a point and the nearest cluster of which that point is not a member \citep{rousseeuw87}.
The average score was calculated for a clustering across all data in the data-set, to evaluate the clustering as a whole and indicate whether isolated groups of sources were identified.
The average score for each cluster was also computed, to measure the similarity and compactness of the sources within a single cluster. 

The silhouette score is one method to indicate the optimal clustering based on cluster isolation. 
Ideally, a maximum silhouette score should be identified from the set of scores for a given colour combination. 
However, with a low number of clusters a high score can be misleading: the clusters in this case appear well isolated due to the large distance between cluster centers. 
If the score peaked at a low number of clusters, the clustering was investigated further to determine if the clustering was optimal.

\section{Analysis}
This section outlines the colour distributions used for clustering and the algorithms' performance. 
Each colour combination was clustered using both K-Means and Mean Shift algorithms, in two and three dimensions.

\subsection{Colour distributions}
\label{sec:col_dist}

The colour distributions for the M83 objects cataloged by \citet{chandar10} showed a broad range of properties, summarized in Table~\ref{tab:colour_stats}.
For broad-band colours, the number of objects with adequate photometry for computing a colour ranged from 
approximately 15~000 for F225W$-$F336W to about 58~000 for $V-I$, with most colours being measured for $3-5\times10^4$ objects.
Fewer objects were detected in the narrow bands:
the number of objects with acceptable narrow-band colours ranged from 8700 (F336W$-$F373N) to about 25~000 (F673N$-$F814W).

\begin{table}
\centering
\caption{ERS catalogue colour distributions}
\label{tab:colour_stats}
\begin{tabular}{lrrrrr}
\hline\hline
Colour & $N$ & mean & median & stdev & IQR\\
& & mag & mag & mag & mag \\
\hline
F225W--F336W &14977 & $-0.07$ & $-0.14$ &$0.43$ &$0.45$\\
F336W--F438W &33523 & $-1.13$ & $-1.21$ &$0.44$ &$0.42$\\
F336W--F555W &33692 & $-1.08$ & $-1.20$ &$0.61$ &$0.58$\\
F336W--F814W &29181 & $-0.79$ & $-0.98$ &$0.96$ &$1.08$\\
F438W--F555W &48660 & $0.15$ & $0.07$ &$0.38$ &$0.35$\\
F438W--F814W &41413 & $0.65$ & $0.43$ &$0.94$ &$1.06$\\
F555W--F814W &57935 & $0.91$ & $0.65$ &$0.98$ &$1.51$\\
F336W--F373N &8675 & $-0.15$ & $-0.25$ &$0.44$ &$0.43$\\
F438W--F487N &13269 & $0.43$ & $0.36$ &$0.34$ &$0.33$\\
F502W--F555W &14644 & $0.01$ & $0.03$ &$0.29$ &$0.23$\\
F657N--F814W &23113 & $-0.81$ & $-0.75$ &$0.88$ &$1.16$\\
F673N-F814W &25185 & $0.33$ & $0.22$ &$0.56$ &$0.70$\\
\hline
\end{tabular}
\end{table}

In addition to including different numbers of objects, the colour distributions also had substantially different shapes.
For most colours the mean colour was larger than the median colour differed, typically by about 0.1~mag.
The standard deviations and inter-quartile ranges (IQRs) were similar (within 20\%) for most of the distributions.
Colours involving a larger wavelength difference between bands had larger standard deviations and IQRs, as might be expected since such colours are more reddening-sensitive.
The smallest and largest colour spreads among the broad-band colours were for $B-V$ and $V-I$, respectively.
For the narrow-band colours, the smallest and largest spreads were for F502N$-V$ and F657-$I$ respectively. 
The figures showing the results of the clustering analysis illustrate many of the individual distributions: 
many can be broadly described as having a blue peak with a fairly sharp cutoff on the blue side, and a more extended red tail. 
This makes sense physically, as the vast majority of objects are expected to be no bluer than a black-body, but spatial variations in reddening will lead to 
a range of colours on the red side of the distribution.

Replacing a broad-band colour with a broad-to-narrow-band colour changes the colour space.
As Tables~\ref{tab:NBcolourcombos} and ~\ref{tab:colour_stats} show, the number of objects with acceptable photometry is reduced.
A colour involving the narrow band should primarily depend on the strength of emission lines in that band.
The effectiveness of clustering will depend on the number of distinct emission-line populations, such as supernova remnants, planetary nebulae, or \ion{H}{ii} regions.
The images shown in figs. 5 and 6 of \citet{blair14} show that M83 supernova remnants appear as resolved, ring-like structures mainly detectable in images taken with narrow-bands.
The object detection for the ERS catalogue was done using the broad-band images and we therefore expect that the catalogue will contain few supernova remnants.
With a maximum size of a few pc, planetary nebulae at the distance of M83 should be unresolved by {\em HST} and would be expected to be included in the catalog.
\ion{H}{ii} regions have a broad range of sizes \citep{hunt09}: the smaller M83 regions should be point sources and therefore be included while larger ones will not.

\begin{figure}
\centering
\includegraphics[width=\columnwidth]{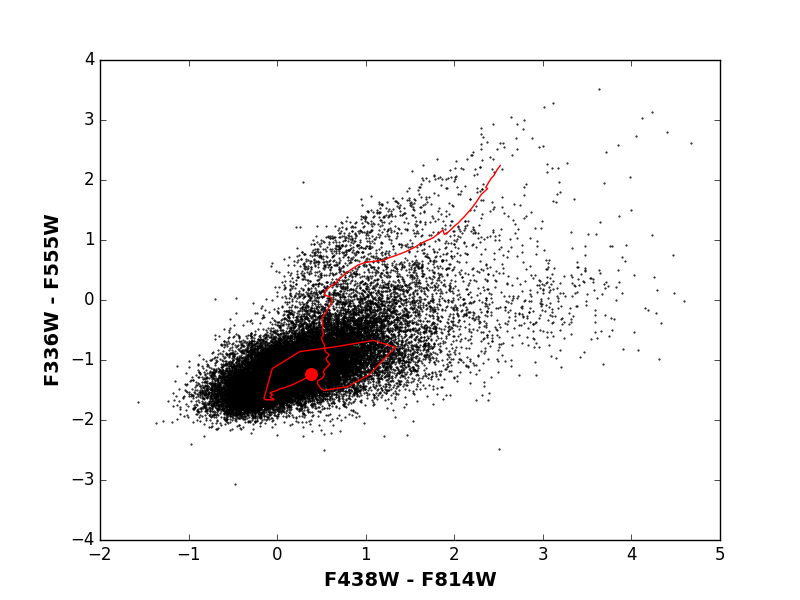}
\caption{Colour-colour distribution for $U - V$ versus $B - I$. 
Compare to Fig.~\ref{fig:model_colour} which shows the same bands in a different combination.
FSPS single stellar population model (${\rm Fe/H}=+0.5$, ages $10^5-10^{10.3}$~yr) is over-plotted as a solid line; the large circle on this line indicates the youngest age.
}
\label{fig:uvbi}
\end{figure}

Many of the two- and three-dimensional colour distributions used for clustering are also shown in the discussion of clustering results.
Some colours are quite well-correlated with each other, particularly if they share a common band (for example, F225W$-U$ and F225W$-V$).
Fig.~\ref{fig:model_colour} illustrates a common feature of both two and three-dimensional colour distributions: 
a concentration of objects with blue colours in both bands and two colour `lobes' at redder colour values. 
This feature was more common in distributions involving broad-band colours; narrow-band colours were more likely to have a single,
more extended `tail' in the direction expected for objects with spectra dominated by emission lines.
In Fig.~\ref{fig:model_colour}, the colours of the blue concentration are consistent with the youngest single stellar population models,
and the older SSP models match one of the red lobes; in other colour combinations the two lobes form the ends of the model age sequence.
As an example of the effects of interchanging bands in forming colours, Fig.~\ref{fig:uvbi} shows the catalogue colours $U - V$ versus $B - I$ --
the same bands as used for Fig.~\ref{fig:model_colour} but in a different combination. 
The colours shown in Fig.~\ref{fig:uvbi} both have a wider wavelength range.
They are more tightly correlated with each other and the two red lobes are less well-separated than in Fig.~\ref{fig:model_colour}.
We therefore focus on the $U - B$ versus $V - I$ clustering in the results discussed below.

\subsection{Algorithm performance}

Figures~\ref{fig:model_colour} and \ref{fig:uvbi} and those below show that the colour-colour distributions of objects in M83 studied here do not contain compact, well-separated clusters.
This is a less-than-ideal situation for detecting groupings in colour space, and the two algorithms used here responded in quite different ways.
The K-Means algorithm generated clusters of roughly comparable size, often by dividing along straight lines in 2-D colour space or planes in 3-D space.
These colour cuts were not necessarily along specific colour axes; in cases where the input colours were correlated, such as in Fig.~\ref{fig:uvbi}, the algorithm
made group divisions perpendicular to the direction of correlation. 
The segmentation did not change dramatically as the number of clusters was increased but rather more finely divided the distribution along the same direction.
In broad-band colours, the K-Means algorithm usually segmented the lobes described above into separate clusters, even for $K=3$ or 4.
In narrow-band colours, $K\geq 5$ was usually required for the emission-line-dominated tail to be selected as a separate cluster.
For most colour combinations, $K=3$ resulted in the largest silhouette score; a plateau in the score as a function of the number of clusters did not occur in all cases.

In contrast to K-Means, the Mean Shift algorithm tended to put most of the objects in a given colour combination into a single large cluster.
The remaining objects were assigned to small clusters (sometimes containing only a single object) located near an edge of the main colour distribution.
The larger (tens to hundreds of objects) `outlier' clusters found by Mean Shift were usually separated from the main body of objects by the same sort of line or plane cut as in K-Means.
While this depended on the bandwidth parameter, in general Mean Shift clustering resulted in a larger number of clusters than considered useful in K-Means.
For the purposes of detecting objects with unusual colours, Mean Shift may be more useful than K-Means, 
but for identifying broad peaks in colour distributions K-Means is superior. 

Two- and three-dimensional colour distributions containing the same colours were not always segmented the same way by the two algorithms.
Similar results were obtained when one of the three dimensional colours had  a very small range or a high correlation with another colour.
In general, the three dimensional colour spaces highlighted the structures visible in two dimensions.
In most narrow band combinations, visual examination showed that two branches of objects separated themselves from the dense centre of the distribution, but neither clustering method in two dimensions was able to identify either branch.
In three dimensions, these branches were more clearly separated from the rest of the distribution, and the clustering methods were able to identify them.
In the discussion that follows, the numeric cluster labels are those arbitrarily assigned by the clustering algorithms; the labels do not imply an ordering.

\begin{figure}
\centering
\includegraphics[width=\linewidth]{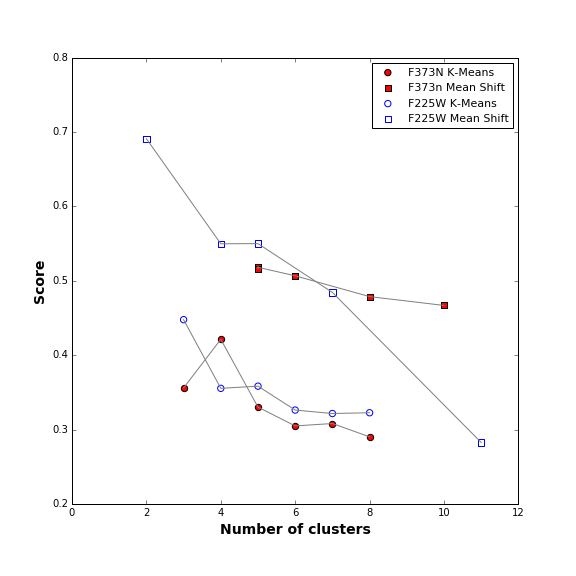}
\caption{Silhouette score as a function of the number of clusters, for two-dimensional clustering in the ${\rm F225W}- U$ and $V - I$ colours (blue open symbols) and for the three dimensional $U - {\rm F373N}$ and $B - I$ combination (red filled symbols). 
Square symbols show the results for Mean Shift clustering and circle symbols for K-Means.
}
\label{fig:sscore}
\end{figure}

Fig.~\ref{fig:sscore} shows the distribution of the silhouette score against the number of clusters for two-dimensional clustering in the $UVW - U$ and $B - I$ colours, and three dimensional clustering in the $U - {\rm F373N}$ and $B - I$ colours.
For the $UVW - U$ and $B - I$ combination using K-Means, the score does not peak in the centre of the distribution.
Instead of selecting the clustering with the highest score, the optimal clustering is found where the relation begins to flatten, between 5 and 6 clusters. 
This clustering is selected because any increase in $K$ after this point does not affect the score.
This means that the algorithm has found the balance between the natural clusters in the distribution and artificially segmenting the data.
For the $U - {\rm F373N}$ and $B - I$ combination, the score for the K-Means method peaks at four clusters, indicating a strong preference for this clustering.

The distribution of silhouette score for the results of the Mean Shift algorithm does not follow the same pattern as for K-Means.
We found that the silhouette score was not as successful at describing the strength of Mean Shift clustering.
Mean Shift often created one large cluster and several very small ones, resulting in a high silhouette score.
For the $UVW - U$ and $B - I$ combination using  Mean Shift, the score decreases linearly with the number of clusters and an optimal clustering cannot be determined from this relation.
For the $U - {\rm F373N}$ and $B - I$ combination, the score is nearly constant with number of clusters.

\begin{figure}
\centering
\includegraphics[width=\linewidth]{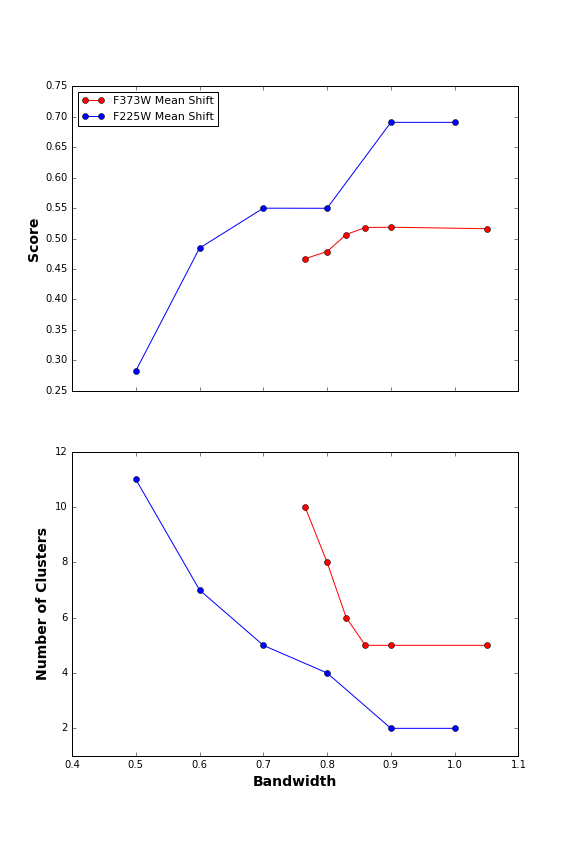}
\caption{Silhouette score and number of clusters as a function of bandwidth for Mean Shift clustering of the three dimensional $U - {\rm F373N}$ and $B - I$ colour combination (red lines), and ${\rm F225W}- U$ and $B - I$ colours (blue lines).}
\label{fig:bwscore}
\end{figure}

In order to determine the optimal Mean Shift clustering, we investigated the relations between the bandwidth and score, and the number of clusters.
Fig.~\ref{fig:bwscore} shows these relations for a three dimensional clustering of the $U - {\rm F373N}$ and $B - I$, and the ${\rm F225W}- U$ and $V - I$ colour combinations. 
In both panels of Fig.~\ref{fig:bwscore} we see that the number of clusters remains at five clusters once the bandwidth passes $0.85$ for $U - {\rm F373N}$ and $B - I$, and only two clusters once the bandwidth passes $0.9$ for the ${\rm F225W}- U$ and $V - I$ combination.
For the  $U - {\rm F373N}$ and $B - I$, the number of clusters and silhouette score converge at the same bandwidth, indicating that the optimal number of clusters had been detected. 

\section{Results}
\label{sect:results}
This section describes the results of the clustering analysis and related to the input colour distributions, with discussions of
the algorithm results in general and some specific examples.

\subsection{Clustering output: broad-band colours}

\begin{figure}
\centering
\includegraphics[width=\columnwidth]{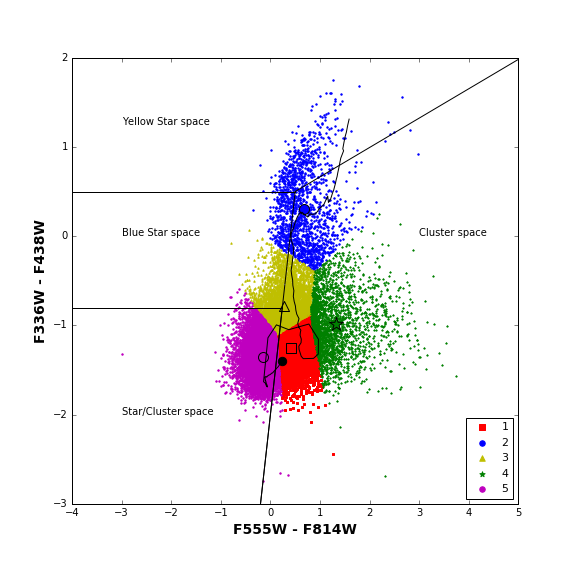}
\caption{$K=5$ K-Means clustering result for two-dimensional $U - B$ and $V - I$ colour distribution. Large symbols show the locations of cluster centres. Overplotted line is the FSPS single stellar population model (${\rm Fe/H}=+0.5$, ages $10^5-10^{10.3}$~yr) with the large circle corresponding to the youngest age.}
\label{fig:broadband_2d_KM5}
\end{figure}

The $UBVI$ bands are very frequently used in {\em HST} studies of both stellar and galaxy populations.
The $U$ and $B$ bands probe the SED peaks for young, hot stars and the metal absorption lines in stellar atmospheres;
the $V$ and $I$ bands probe cooler stars and highly-reddened populations.
Roughly 29~000 objects in the M83 field have reliable photometry in all four bands, with $UB$ detections being the limiting factor.
Clustering the $U - B$ and $V - I$ colours in two dimensions with the K-Means method identified five clusters as the optimal number,
via a plateau in the silhouette score.
Fig.~\ref{fig:broadband_2d_KM5} shows the cluster assignments and we discuss them below in descending order of size.

The clusters identified by K-Means in $UBVI$ vary in both number of objects and colour spread.
The largest clusters, \#1 and \#5, are also the most compact in colour space and correspond to the bluest colours in both bands.
Both are located in the `star/cluster' region identified in the same colours by \citet{chandar10}.
Cluster \#1 with its slightly redder $V-I$ colours also overlaps with the loop in colour space predicted to be followed by simple stellar populations with ages between $10^7$ and $10^8$~yr (see Fig.~\ref{fig:model_colour}).
Cluster \#3 is redder in $U-B$ than clusters 1 and 5 and corresponds to the `blue star' region identified by \citet{chandar10}.
Comparison with figure~4 of \citet{kim12} suggests that members of all three groups are likely to be bright main-sequence stars with varying degrees of reddening.
Cluster \#4 spans the same range in $U-B$ as \#1, \#3 and \#5, but is the reddest in $V-I$, falling within the `cluster' colours identified by \citet{chandar10},
and the `3\%' region ($V - I >1.2$) shown by \citet{kim12}. 
Most of the supergiant star candidates identified by \citet{williams15} lie in this region.
\citet{kim12} suggested that many of the colours in this region of the diagram were the result of incorrect matching between different bands. 
(This should not be a problem in our analysis, since the catalogue used here was constructed differently; we have also verified that sources with these colours do not all have high photometric uncertainties.)
Cluster \#2, the reddest in $U-B$, was identified as a separate cluster even with $K<5$.
This grouping spans the `yellow star' and [star] `cluster' colours identified by \citet{chandar10} and corresponds to the oldest stellar populations.
\citet{kim12} identified this colour region as belonging to non-main sequence stars, whose $UBVI$ colours depend on age as well as reddening.

\begin{figure}
\centering
\includegraphics[width=\columnwidth]{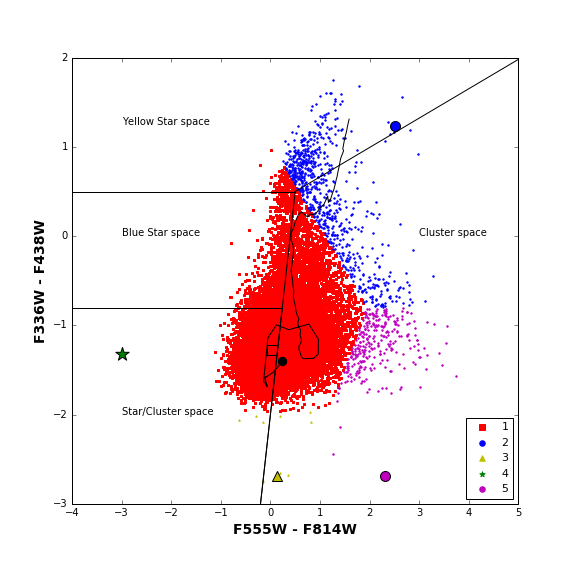}
\caption{Colour-Colour distribution of the $U - B$ and $V - I$ colours, clustered using Mean Shift with $h=0.6$ producing five clusters. Large symbols show the locations of cluster centres. Overplotted line is the FSPS single stellar population model (${\rm Fe/H}=+0.5$, ages $10^5-10^{10.3}$~yr) with the large circle corresponding to the youngest age.}
\label{fig:broadband_2d_MS5}
\end{figure}

Clustering the $UBVI$ distribution with Mean Shift in two dimensions produced very different results from K-Means (Fig.~\ref{fig:broadband_2d_MS5}).
The bandwidth value $h=0.6$ marked the point at which the number of clusters no longer increased with bandwidth; this value produced five clusters.
(Five-cluster segmentations with different bandwidth values were not substantially different.)

However, clusters \#3 and \#5 contain only three objects in total  whereas K-Means distributed objects far more evenly between clusters.
There is little agreement in cluster assignment between Mean Shift and K-Means, except for K-Means cluster 2, the reddest in $U-B$.
Despite this apparent success in identifying different features of the distribution, the K-Means result had lower silhouette score (0.36) than  Mean Shift (0.40). 
The greater colour separation of the two very small clusters found by Mean Shift increases its silhouette score.
In this case the silhouette score as a measure of clustering effectiveness seems to be inadequate.

\begin{figure*}
\includegraphics[width=0.45\textwidth]{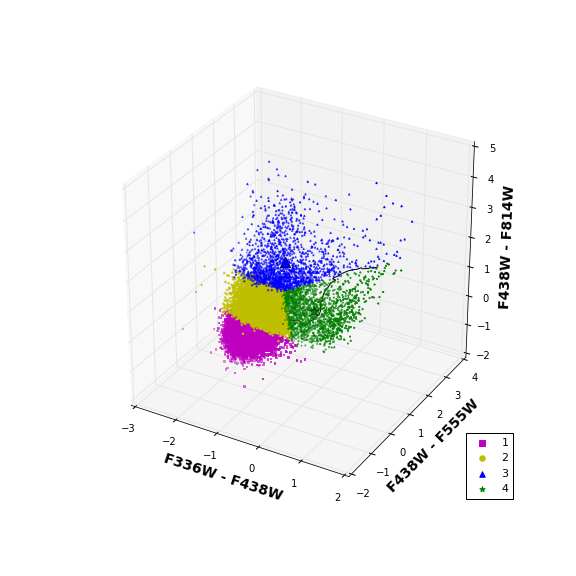}
\includegraphics[width=0.45\textwidth]{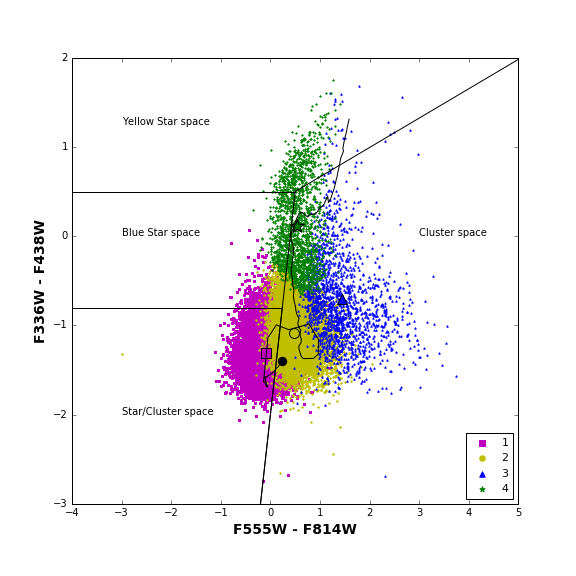}

\includegraphics[width=0.45\textwidth]{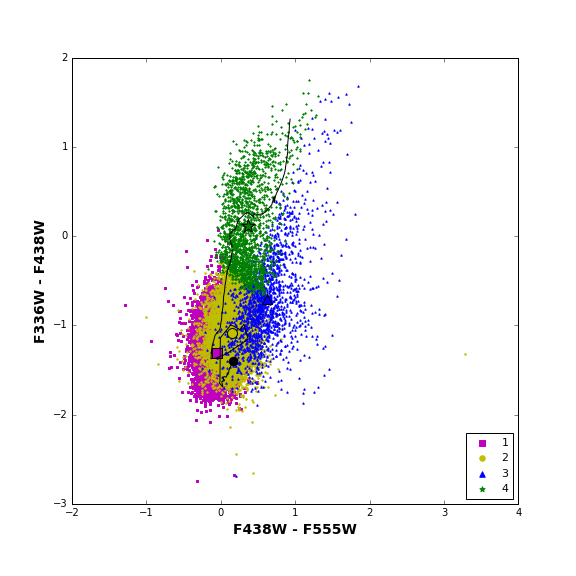}
\hfill
\includegraphics[width=0.45\textwidth]{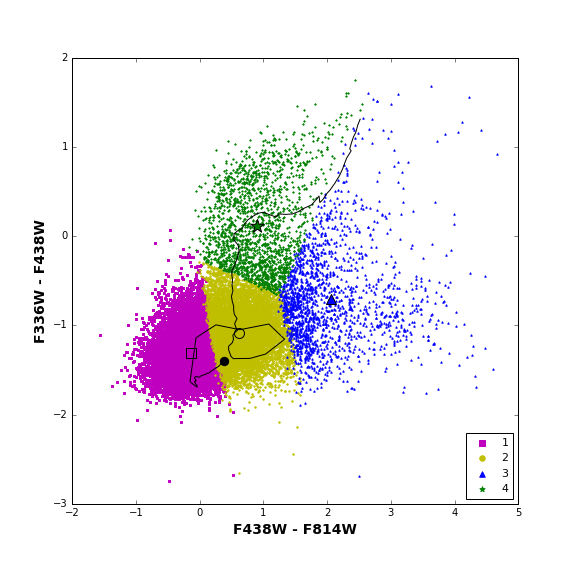}

\caption{K-Means $K=4$ segmentation for three-dimensional clustering of the $U - B$, $B - V$, and $B - I$ colours. Large symbols show the locations of cluster centres. Overplotted line is the FSPS single stellar population model (${\rm Fe/H}=+0.5$, ages $10^5-10^{10.3}$~yr) with the large circle corresponding to the youngest age.}
\label{fig:broadband_3d_KM}
\end{figure*}

To examine the $UBVI$ distribution in three dimensions, the $U - B$, $B - V$, and $B - I$ colours were used.
The colours in this combination had a larger range than other possible combinations, and it was expected that they would lead to more separation of branches  in the colour space. 
K-Means clustering in three dimensions in the $UBVI$ bands produces a slightly different result from the two-dimensional clustering:
the peak silhouette score occurs for four clusters instead of five. 
Fig.~\ref{fig:broadband_3d_KM} shows the assignments in the three-dimensional colour space and projections onto several two-dimensional spaces.
In three-dimensional colour space, a main concentration and two branches are evident (top left panel) although 
their separation in the colour space projections (remaining panels)  is less clear than in the two-dimensional case.
The separation between clusters roughly follows planes in the three-dimensional space with no clear decrease in density of points at the cluster boundaries.
The clustering identifies two large groups of blue objects (clusters \#1/2 in 3-D; roughly clusters \#5/3/1 in 2-D),
one smaller group of red objects (\#3 in 3-D, \#4 in 2-D) and an intermediate-colour group, the smallest (\#4 in 3-D, \#2 in 2-D).

\begin{figure*}
\centering
\includegraphics[width=0.45\textwidth]{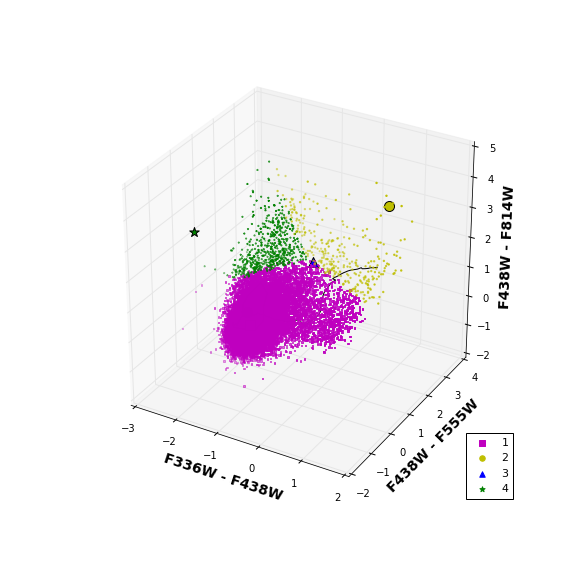}
\includegraphics[width=0.45\textwidth]{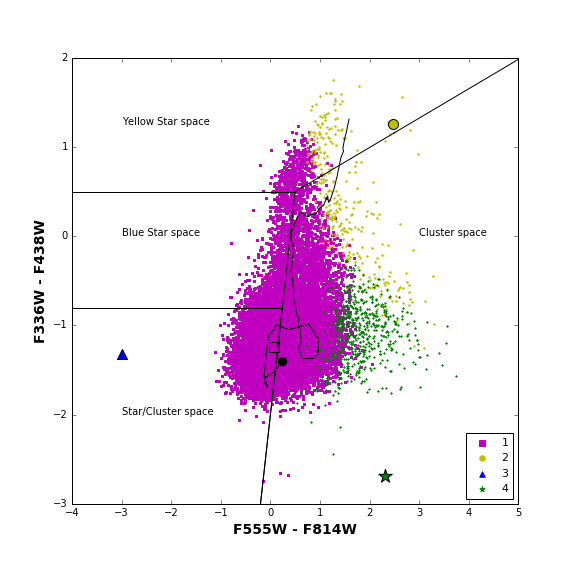}
\caption{Mean Shift $h=0.75$ segmentation for three-dimensional clustering of $U - B$, $B - V$, and $B - I$ colours. Large symbols show the locations of cluster centres. Overplotted line is the FSPS single stellar population model (${\rm Fe/H}=+0.5$, ages $10^5-10^{10.3}$~yr) with the large circle corresponding to the youngest age.}
\label{fig:broadband_3d_MS4}
\end{figure*}

The three dimensional $UBVI$ clustering with Mean Shift produced larger clusters than typical for this algorithm. 
Most bandwidth settings produced 6 clusters which were not well-separated, but a bandwidth $h=0.75$ produced 4 clusters and the highest silhouette score. 
The left panel of Fig.~\ref{fig:broadband_3d_MS4} shows the segmentation, while the right panel shows the projection into the original space, where the cluster locations are easier to identify.
The algorithm identified two groups of objects (clusters \#2 and \#4) which are red in $V - I$ but differ in $U - B$ colours.  
These two groups combined only comprise 3\% of the objects but are also identified at other bandwidth values.
Cluster \#2 (red in both $U-B$ and $V-I$) contains brighter objects more evenly spread across the galaxy while cluster \#4
(blue in $U-B$ but red in $V-I$) contains fainter objects more tightly associated with the spiral arms.
Cluster \#3 as identified by Mean Shift includes only a single object, whose catalogue entry shows an extremely blue $V-I$ colour and an extremely red $B-V$ colour, likely indicating a problem with the F555W photometry.
The remainder of the objects are contained in a single cluster, highlighting Mean Shift's ability to find outliers in the distribution at the expense of segmenting more dense regions of colour space;
for example, it does not select as a separate group the large branch of objects that are red in $U - B$.

In summary, M83 objects detected in all of the $UBVI$ bands could be separated into groups using either the K-Means or Mean Shift algorithms.
Constructing three colours, rather than two, from the four bands did not result in a strikingly different segmentation.
The $UBVI$ colour groups identified by $K$-Means can be roughly identified with bright main-sequence stars (the majority of objects), star cluster candidates, and non-main sequence stars.
As the colour groups are contiguous, these identifications are unlikely to be definitive.
Lacking {\em a priori} classifications for most objects, we cannot directly evaluate the accuracy of clustering-based separation into different classes. 
A check on the usefulness of clustering in broad-band colour space can be made by examining the magnitude distributions of the different groups.
The distribution of $V$ magnitudes for objects within the clusters identified in either the two- or three-dimensional K-Means analysis extends to the catalogue limit,
except for clusters \#2 in 2-D and \#4 in 3-D, whose faintest members were nearly 2 magnitudes above the limit.
The location of these objects in colour-magnitude space suggests that they are good candidates to be M83 star clusters.
Comparison with published classifications shows that, compared to the other K-Means groups, these groups do contain a higher fraction of objects classified as star clusters.
However, the fraction of objects with classifications is very small and changing the classification of a handful objects would change this result.
The reddest groups of objects (\#4 in 2-D and \#3 in 3-D) are candidates to be background galaxies,
but existing published classifications are insufficient to test this.

\subsection{Clustering output: narrow-band colours}

\begin{figure*}
\centering
\includegraphics[width=0.45\linewidth]{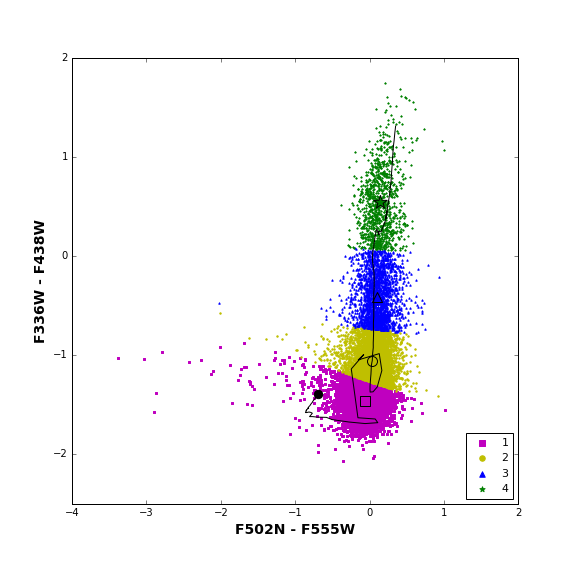}
\includegraphics[width=0.45\linewidth]{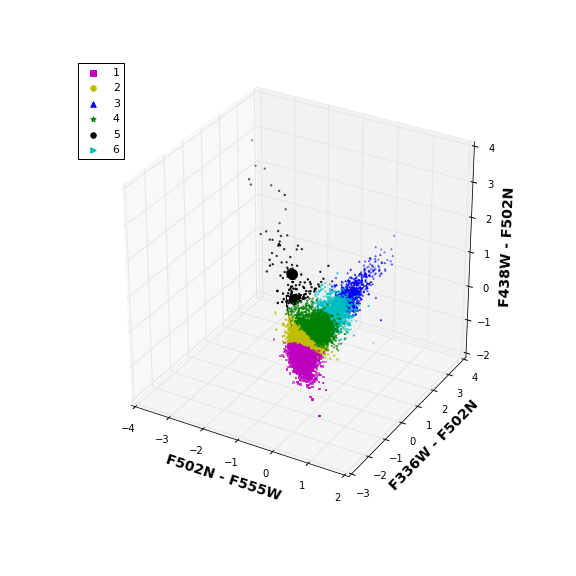}
\caption{Colour distributions formed from $UBV$ and F502N.
  Left: result of two-dimensional clustering using K-Means with $K=4$.
  Right: result of three-dimensional clustering using K-Means with $K=6$. Large symbols show the locations of cluster centres. Overplotted line is the FSPS single stellar population model (${\rm Fe/H}=+0.5$, ages $10^5-10^{10.3}$~yr) with the large circle corresponding to the youngest age.}
\label{fig:oiii_IV_KM}
\end{figure*}

As discussed above, K-Means clustering for many of the narrow-band colour combinations resulted in segmentation primarily
along {\em broad}-band colour axes.
The K-Means result from the $UBV$ and F502N bands is shown as an example in Fig.~\ref{fig:oiii_IV_KM}:
in two dimensions the optimal number of clusters was found to be four and in three dimensions, six.
However, Fig.~\ref{fig:oiii_IV_KM} shows that the clusters are separated primarily along the $U-B$ colour axis:
the only distinct set of objects is dominated by emission lines and therefore blue in ${\rm F502N}-V$.
This group could have been selected based on the F502N and $V$ photometry alone.
Results for F487N, F657N, and F673N were similar.

\begin{figure}
\centering
\includegraphics[width=\columnwidth]{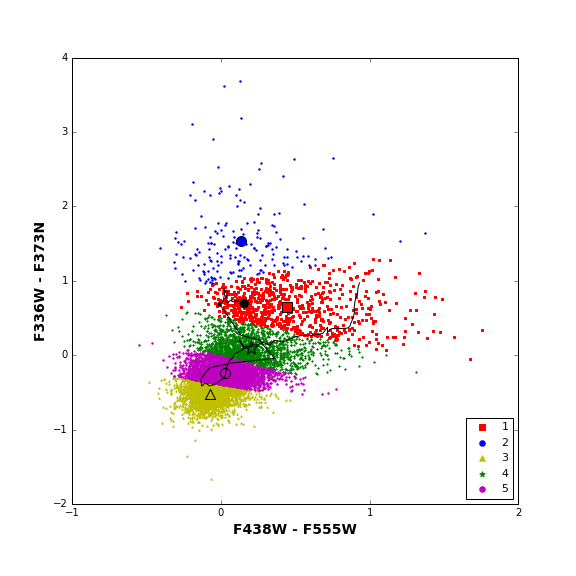}
\caption{K-Means segmentation of $U-{\rm F373N}$ and $B-V$ colours, with $K=5$. Large symbols show the locations of cluster centres. Overplotted line is the FSPS single stellar population model (${\rm Fe/H}=+0.5$, ages $10^5-10^{10.3}$~yr) with the large circle corresponding to the youngest age.}
\label{fig:UOII_2d_KM5}
\end{figure}

\begin{figure*}
\centering
\includegraphics[width=0.45\textwidth]{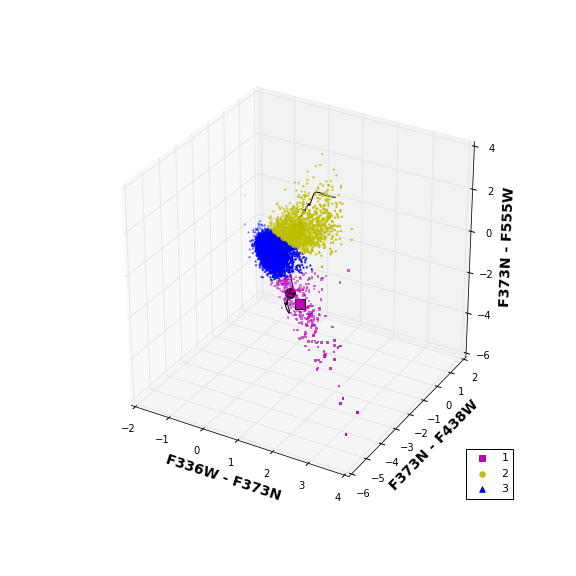}
\includegraphics[width=0.45\textwidth]{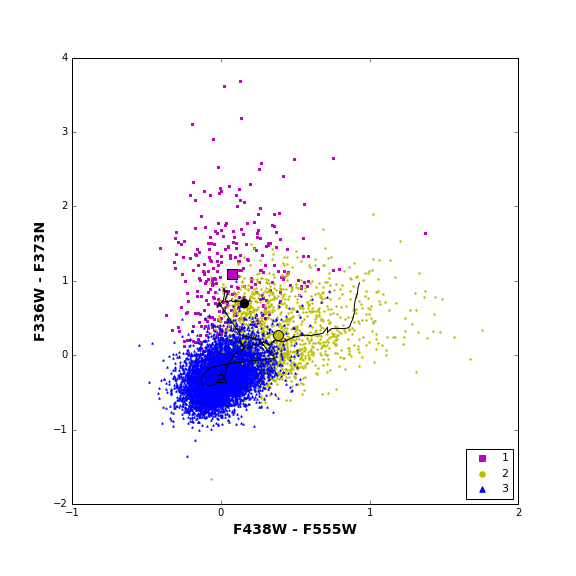}
\caption{K-Means $K=3$ segmentation of three-dimensional colour distribution $U-{\rm F373N}$, ${\rm F373N}-B$, and ${\rm F373N}-V$ (left) and the same distribution in projection (right). Large symbols show the locations of cluster centres. Overplotted line is the FSPS single stellar population model (${\rm Fe/H}=+0.5$, ages $10^5-10^{10.3}$~yr) with the large circle corresponding to the youngest age.}
\label{fig:UOII_3d_KM}
\end{figure*}

The colour space formed by the $U$, F373N, $B$ and $V$ bands produced different clustering results from the other narrow bands.
In two dimensions, $K$-Means produced a meaningful segmentation, with 5 clusters at the elbow of the silhouette score versus $K$ plot.
The segmentation (Fig.~\ref{fig:UOII_2d_KM5}) is primarily along the $U - {\rm F373N}$ axis; 
the $B-V$ colour has little effect on the clustering in this space.
The simple stellar population models that both the youngest and oldest populations will have red $U - {\rm F373N}$ colours,
with bluer colours belonging to the intermediate age ($10^7-10^8$~yr) loop. 
Most of the objects in this sample are in the bluer groups.
The K-Means result also includes a group with very red colours indicative of strong \ion{O}{iii} line emission not predicted by simple stellar population models.
These objects could be planetary nebulae; the high metallicity of M83 is expected to result in weak \ion{O}{iii} emission by \ion{H}{ii} regions \citep{blair14}.

Fig.~\ref{fig:UOII_3d_KM} shows the three-dimensional distribution in this band combination with the K-Means clustering result.
The optimal clustering in this case is for three groups, rather than five. 
The distribution here shows two fairly clear branches, not driven by emission line dominance but by the ${\rm F373N}- B$ and ${\rm F373N}- V$ colours.
In this case the three clusters form an age sequence: according to the simple stellar population model predictions, cluster \#1 is young, cluster \#3 is intermediate-age, and cluseter \#2 is old. 
The extreme end of group 1 is emission-line dominated.
This population is not picked out as a separate cluster by K-Means until $K$ reaches 7, although Mean Shift selects it as one of four clusters.
The differences between the two- and three-dimensional results for the F373N band, while atypical of our results in general, indicate that selecting colour spaces for clustering requires careful thought and considerable experimentation.

\section{Discussion and Conclusions}
The classic colour-magnitude diagram has endured in observational astronomy because of its simplicity and ability to be translated into physical parameters.
Two-colour diagrams from three- or four-band photometry can be more difficult to interpret but can also have broader utility, 
from measuring reddening of main-sequence stars to identifying active galactic nuclei.  
Three-colour diagrams in three dimensions are relatively rarely used but are becoming more common.  
Understanding multi-colour distributions will be of increasing importance as larger surveys become the norm.
Colour distributions of point sources within nearby galaxies are complex, due to the wide range of age, metallicity, and reddening found within galaxies. 
Identifying specific classes of sources often relies on spatial (location and morphology) and luminosity information as well as colour. 
Our analysis found that broad-band colours of point sources in M83 did not divide neatly into isolated colour groups.

The segmentation appeared to be more robust in three dimensional colour combinations compared to two dimensions and we attribute
this to the additional information content of the higher dimensional space.
Correcting for the spatially-varying effects of reddening internal to M83 \citep[e.g., with a method like that used by][]{dalcanton15} might have improved the separation of groups; however this comes at the cost of making strong assumptions about the content of the catalogue.

The two algorithms we tested for clustering the colours of point sources in M83 showed quite different behaviour in segmenting the colour distributions.
The K-Means algorithm divided colour distributions relatively evenly along lines or planes in colour space.
Although choosing the number of clusters can be a difficulty in using this algorithm, we found that the clusters identified changed smoothly between different values of $K$.
This is likely why the silhouette score did not vary substantially between $K$ values.
In contrast, Mean Shift was able to create clusters of uneven sizes, particularly when the distribution contained branches of objects spanning a large colour range.
The Mean Shift results were often sensitive to the bandwidth parameter; the elbow in the relation between bandwidth and silhouette score could sometimes be used to identify a reasonable clustering.
In this case, the algorithm would pick out small groups of objects on the edges of the main distribution.
We recommend that searches for small groups of outliers in colour space consider using Mean Shift.

For identifying larger groups of objects with similar colours,  we found K-Means to be more effective.
The broad band colour combination that was most effective at identifying different colour classes of objects was $U - B$ and $V - I$;
in both two and three dimensions this combination was more effective than $U - V$ and $B - I$. 
It showed a clear branch of objects red in  $U - B$ which are good candidates to be M83 star clusters; the $K$-means algorithm was able to identify this group as distinct.
Comparison with previous work on the stellar content of M83 allowed us to make some general conclusions about the nature of the colour groups, 
but the difficulty in matching the catalogue with previously-categorized objects prevented detailed comparisons.
All of the narrow band colour combinations showed branches of emission-line dominated objects.
K-Means identified these as separate groups if the number of clusters $K\geq 5$ but the Mean Shift algorithm generally identified only the extreme members of these groups.

In this work we have restricted our analysis to colours that can be generated with observations in four bandpasses. 
The multi-band nature of the M83 dataset allows further exploration in higher-dimensional clustering, which we intend to pursue in future work.
The results of the present analysis allow us to make some recommendations for bandpass selection for UV-visible imaging surveys of nearby galaxies. 
Of course, the target, specific science goals, and instrumentation used for an observation will likely play a major role as well, but for general-purpose studies of a galaxy's point source population, we suggest that bandpasses be chosen to:
\begin{description} 
\item cover as wide a wavelength range as possible 
\item include $B$ or $V$, but not both, if only 3 broad bands are used
\item include a limited number of narrow bands, ideally H~$\alpha$.
\end{description}
Matching observation depth across a wide wavelength range can be difficult if instrument sensitivity also varies with wavelength, 
but is crucial for a full characterization.
Clustering techniques such as K-Means and Mean Shift show promise in identifying general trends and small populations of outliers, although the continuous nature of colour distributions means that they will likely be used to suggest additional analyses rather than producing definitive identifications.
Analysis of data from future multi-band sky surveys with both ground- and space-based facilities will benefit from further exploration of these techniques.

\section*{Acknowledgments}

The authors acknowledge financial support from the Natural Science and Engineering Research Council (NSERC) of Canada. 
We thank S. Lianou for helpful comments on the manuscript.
This research has made use of the NASA/IPAC Extragalactic Database (NED) which is operated by the Jet Propulsion Laboratory,
California Institute of Technology, under contract with the National Aeronautics and Space Administration. 
This research has made use of the SIMBAD database, operated at CDS, Strasbourg, France.
We acknowledge the efforts of the WFC3 Science Oversight Committee in conducting the Early Release Science program.

\bibliographystyle{mnras}
\bibliography{m83_refs}

\appendix
\section{Published Catalogues}
\label{appendix:pubcat}

One measure of success for an unsupervised clustering process is the degree to which the cluster correspond to previously-identified classes of objects. 
We planned to do this by  
compiling a `published  catalogue' combining the contents of the objects near M83 listed in the NASA Extragalactic Database (NED) and the Set of Identifications, Measurements and Bibliography for Astronomical Data \citep[SIMBAD][]{wenger2000}. 
To these we added  catalogues of Wolf-Rayet stars \citep{kim12} and
red supergiant candidates \citep{williams15} which did not appear in either database.
NED's focus as an extragalactic database and SIMBAD's focus on Galactic objects mean that their contents overlap but are not identical, 
and this is true of the area surrounding M83. 
A $3\farcm3$ radius region around the coordinates centered at 204.26761$^{\circ}$, $-29.839939^{\circ}$ contains 1~553 NED objects and 1~772 SIMBAD objects, of which 1~220 are matched with each other at 1~arcsec tolerance.
Although the two services use slightly different naming conventions, with human inspection the matches are generally recognizable as referring  to the same object. 
Interestingly, the databases do not always report the same object type even when the names are identical.
The differences are reasonable in some cases (a supernova remnant can also be an X--ray source, for example), but not others (e.g. CXOU J133703.0-294945 is reported as a supernova remnant by SIMBAD and an \ion{H}{ii}  region by NED).
Objects which appeared in one database but not the other were primarily from recent work \citep[e.g.][]{long2014}, from older studies likely superseded by newer ones  \citep[e.g.][]{larsen1999}, or from studies in which only coordinates relative to the galaxy centre were given \citep{rumstay83,dvpd83}.
Our final combined  catalogue had 2~425 objects of which approximately 750 are in the region covered by the ERS  catalogue.
Just under half (340) of the 750 are star clusters, with  X--ray sources, supernova remnants, and \ion{H}{ii} regions comprising about 90 objects each.

While compiling the published  catalogue was relatively straightforward, matching its entries to those in the ERS  catalogue was surprisingly difficult.
The nearly 100-fold difference in object density between the two  catalogues was a clue that matching based on sky position alone was unlikely to be successful.
Nearly every entry in the published  catalogue had an ERS  catalogue object within 1\arcsec\ (mean distance between matches was 0\farcs26).
However, visual inspection of the `whitelight' (combined UBVI) image used to generate the ERS  catalogue showed that most of the matches were to very faint sources which would not likely have been detectable in the data sets used to make the published  catalogues. 
By comparing the ERS  catalogue positions to single-filter WFC3 image mosaics available through the Hubble Legacy Archive, we found that the ERS  catalogue positions had a small astrometric offset: they were approximately 0\farcs6 too large in right ascension and 1\farcs0 too large in declination. 
Correcting this offset led to good matches between astrometric standard stars (e.g.\ 2MASS, UCAC2) with bright stars in the WFC3 images, but other published objects were still overwhelmingly matched with very faint objects in the ERS  catalogue.
While it might have been possible to improve the matching using magnitude or colour information, this could also have biased a comparison to the clustering results.
We concluded that the precision and accuracy of published positions for individual objects within M83 was not sufficient for confident matching based on positions alone, and so decided not to pursue the comparison between published  catalogues and unsupervised clustering further.

\bsp
\end{document}